\documentclass[conference]{IEEEtran}

\usepackage[letterpaper, left=0.680in, right=0.680in, bottom=1.049in, top=0.71in]{geometry}

\usepackage{ifpdf}
\ifCLASSINFOpdf
\usepackage[pdftex]{graphicx}
\graphicspath{{./Figures/}}
\DeclareGraphicsExtensions{.pdf,.jpeg,.png}
\else
\usepackage[dvips]{graphicx}
\DeclareGraphicsExtensions{.pdf}
\usepackage[caption=false, font=footnotesize]{subfig}
\fi

\usepackage{tabularx, booktabs}
\usepackage{amssymb}
\usepackage{amsthm} 
\usepackage[ruled,vlined,linesnumbered]{algorithm2e}
\usepackage{aligned-overset} 
\usepackage{array}
\usepackage{balance}
\usepackage{cite}
\usepackage{color}
\usepackage{epstopdf}
\usepackage{enumitem} 
\usepackage{mathtools} 
\usepackage{makecell} 
\usepackage{multicol} 
\usepackage{multirow} 
\usepackage[caption=false,font=small]{subfig}
\usepackage{textcomp}
\usepackage{stfloats}
\usepackage{url}
\usepackage{verbatim}
\usepackage{graphicx}
\usepackage{url}
\usepackage{float}
\usepackage{xcolor}

\definecolor{cyan}{RGB}{0, 102, 204} %
\definecolor{red}{RGB}{204, 0, 0} %
\definecolor{green}{RGB}{5,107,68} %
\definecolor{purple}{RGB}{102, 51, 153} %
\definecolor{blue}{rgb}{0.06, 0.2, 0.65}

\usepackage{hyperref}
\hypersetup{colorlinks=true,linkcolor=red,urlcolor=blue, citecolor=blue}

\newtheorem{remark}{Remark}

\begin{document}

\title{Neural Network-based Information-Theoretic Transceivers for High-Order Modulation Schemes}

\author{
    \IEEEauthorblockN{Ngoc~Long~Pham~and~Tri~Nhu~Do}
    \IEEEauthorblockA{ Telecom Neural Detection Lab, Polytechnique Montr\'{e}al, Montreal, QC, Canada. }
    \IEEEauthorblockA{ Emails: phamngoclong3005@gmail.com, tri-nhu.do@polymtl.ca }
	}
\maketitle

\begin{abstract}
Neural network (NN)-based end-to-end (E2E) communication systems, in which each system component may consist of a portion of a neural network, have been investigated as potential tools for developing artificial intelligence (AI)-native E2E systems. In this paper, we propose an NN-based bitwise receiver that improves computational efficiency while maintaining performance comparable to baseline demappers. Building on this foundation, we introduce a novel symbol-wise autoencoder (AE)-based E2E system that jointly optimizes the transmitter and receiver at the physical layer. We evaluate the proposed NN-based receiver using bit-error rate (BER) analysis to confirm that the numerical BER achieved by NN-based receivers or transceivers is accurate. Results demonstrate that the AE-based system outperforms baseline architectures, particularly for higher-order modulation schemes. We further show that the training signal-to-noise ratio (SNR) significantly affects the performance of the systems when inference is conducted at different SNR levels.
\end{abstract}

\begin{IEEEkeywords}
AI/ML, End-to-End Learning, Log-likelihood Ratio, CNN, AE, Bit-error-rate (BER)
\end{IEEEkeywords}

\section{Introduction}

As sixth-generation (6G) wireless technologies advance, neural network (NN)-based E2E communication systems offer a transformative approach to optimizing spectral efficiency, reducing latency, and enhancing adaptability across diverse channel conditions. These systems model each component as an NN, enabling joint optimization of the transceiver. In contrast, traditional systems rely on probabilistic models, with the log-likelihood ratio (LLR) as the primary method for information extraction at the receiver \cite{o2017introduction}.
Soft decision decoding (SDD) leverages probabilistic information to enhance the accuracy of symbol or bit detection from noisy received signals. A priori probability (APP)-based SDD computes the complete set of a posteriori probabilities for all possible symbols, providing soft outputs, such as LLRs, for iterative decoding. Conversely, maximum a posteriori (MAP) decoding selects the most likely symbol by maximizing the a posteriori probability, typically yielding a hard decision. NN-based classification, using one-hot encoded symbol indices, can approximate MAP decoding by learning to output probabilities that reflect a posteriori probabilities, offering a flexible and trainable alternative for high-order modulation schemes.

Recent research on NN-based systems in the physical layer has demonstrated their ability to learn robust input-output mappings under non-ideal channel conditions, achieving performance comparable to traditional decoding methods, particularly for lower modulation orders \cite{o2017introduction,dorner2017deep,2019cammerer}. In \cite{2020Hoydis}, a trainable NN-based soft demodulator, termed LLRNet, is proposed, which outperforms conventional methods. A deep learning (DL)-based receiver, designed to replace the entire signal reception and recovery process, is introduced in \cite{zheng2020deepreceiver}. Additionally, a fully convolutional neural network (CNN)-based demodulator, capable of performing both hard and soft demodulation, is presented in \cite{zheng2022demodnet}.
Traditional communication systems rely on suboptimal designs, where individual components are optimized independently, failing to ensure global optimality. To address this limitation, AE-based architectures were introduced in \cite{Oshea2017}, modeling multiple communication blocks at the transmitter and receiver as a unified E2E system using interconnected deep neural networks (DNNs) trained jointly. In \cite{zhu2019joint}, a convolutional AE is proposed for joint optimization of the transmitter and receiver, demonstrating superior BER performance under fading channels compared to traditional methods. Furthermore, \cite{2019cammerer} presents a trainable point-to-point communication system incorporating a fully differentiable neural iterative demapping and decoding structure.
Despite these advancements, \textit{a critical research gap} persists: high-order modulation with deterministic constellations and recent E2E AE architectures often entail high computational complexity, which impacts training efficiency. Moreover, most systems are trained under fixed conditions, optimizing solely for BER, which limits their adaptability to varying channel conditions.

\textit{In this paper}, we focus on NN-based E2E information-theoretic transceivers for high-order modulation, designing DL approaches to approximate MAP decoding.
The main contributions\footnote{The code repository can be accessed at: \url{https://github.com/TND-Lab/Neural-Network-based-Information-Theoretic-Transceivers}} are as follows:
\begin{itemize}
\item To address high-order modulation, we propose an NN-based demapper to tackle the bit-wise LLR regression problem and an AE-CNN E2E transceiver to address the symbol-wise classification problem.
\item We validate the BER performance of the proposed NN-based demapper and transceiver against theoretical BER to confirm the accuracy of our solutions.
\item We demonstrate that SNR used during training significantly affects performance, leading to varying outcomes in the inference phase.
\item Our results demonstrate that NN-based solutions outperform conventional methods in low SNR regimes with moderate modulation orders.
\end{itemize}

\section{System Description and Problem Formulation}

\subsection{Information-Theoretic Communication System with SDD}

Consider an information-theoretic communication framework. Let 
$\vec{b} = (b_1, b_2, \ldots, b_n)$ be a sequence of $n$ input bits, where each $b_i \in \{0,1\}$. 

\subsubsection*{Transmitter}

At the transmitter side, for \textit{bit grouping}, the input bitstream $\vec{b}$ is partitioned into consecutive blocks of length $k$. The sequence of blocks can be written as $\vec{c} = (\vec{c}_1, \vec{c}_2, \ldots, \vec{c}_i, \ldots, \vec{c}_{n/k})$. Assuming, without loss of generality, that $n$ is divisible by $k$, such that $n \mod k = 0$, each block is defined as
\begin{align} \label{eq_bit_block}
\vec{c}_i = (b_{(i-1)k+1}, b_{(i-1)k+2}, \ldots, b_{ik}) \in \{0,1\}^k
\end{align}
for $i = 1, 2, \ldots, n/k$. Each $\vec{c}_i$ is thus a $k$-bit information block. Note that $\vec{c} = [\vec{c}_1, \ldots, \vec{c}_i, \ldots, \vec{c}_{n/k}] \equiv \vec{b}$.

Next, for \textit{symbol assignment}, since each block $\vec{c}_i$ contains $k$ bits, there are $M = 2^k$ possible $k$-bit messages. A mapping function:
\begin{align}\label{eq_gray_symbol}
f: \{0,1\}^k \longrightarrow \mathcal{S} = \{s_1, s_2, \ldots, s_m, \ldots, s_M\},
\end{align}
assigns each unique $k$-bit message to one of the $M$ distinct symbols, where $m \in \{1, 2, \ldots, M\}$. In our work, these symbols are labeled using Gray coding to reduce the probability of bit errors for adjacent constellation points. Each $s_m$\footnote{Note that we do not denote $s_m$ as a vector; rather, we denote $s_m$ as a symbol in a codebook, e.g., $s_1 = (0000)$ in a Gray codebook.} is a symbol in the Gray codebook.

At the final modulation step, for \textit{constellation mapping}, each symbol $s_m$ is mapped to a complex-valued constellation point $x_m$ in $\mathbb{C}$. Thus, the overall modulator
\begin{align}\label{eq_complex_symbol}
\mathcal{S} \longrightarrow \mathcal{X} = \{x_1, \ldots, x_m, \ldots, x_M\} \subset \mathbb{C},
\end{align}
maps each $k$-bit message $\vec{c}_i$ to the corresponding complex-valued point $x_m$ in the constellation $\mathcal{X}$. 
The output of the modulator is a complex-valued sequence:
\begin{align}\label{eq_sym_vec}
\vec{x} = (x^{(1)}, x^{(2)}, \ldots, x^{(n/k)}),
\end{align}
where each $x^{(i)} \in \mathcal{X}$ is the complex symbol corresponding to the information block $\vec{c}_i$, which is subsequently transmitted over the physical channel.

Further considering the \textit{normalization} of modulated symbols to ensure unit average energy, $\vec{x}$ is normalized as follows $\vec{x}_{\text{norm}} = \alpha \vec{x}$, where the scaling factor $\alpha$ satisfies
\begin{align}\label{eq_alpha_norm}
|\alpha|^2 E_{\text{avg}} = 1, \quad \text{where} \quad E_{\text{avg}} = \textstyle \frac{1}{n/k} \sum_{i=1}^{n/k} |x^{(i)}|^2.
\end{align}
Thus, $\alpha = 1 / \sqrt{E_{\text{avg}}}$.
For a uniform $M$-ary constellation, the average energy of $\mathcal{X}$ is computed as $E_{\mathcal{X}} = \textstyle \frac{1}{M} \sum_{m=1}^M |x_m|^2$.
If the constellation is designed with equal probability for each point, then $E_{\text{avg}} = E_{\mathcal{X}}$, so
$\vec{x}_{\text{norm}} = (1/\sqrt{E_{\mathcal{X}}} )\vec{x}$.
Hence, the normalized transmitted symbol sequence is:
\begin{align}\label{eq_vec_x_norm}
\vec{x}_{\text{norm}} = (x^{(1)}_{\text{norm}}, x^{(2)}_{\text{norm}}, \ldots, x^{(n/k)}_{\text{norm}}),
\end{align}
where each $x^{(i)}_{\text{norm}} \in \mathcal{X}_{\text{norm}}$.

\subsubsection*{Probabilistic Channel Modeling}

In this work, we characterize the communication channel through its probabilistic description, defined by the conditional probability density function $p(\vec{y} \mid \vec{x})$, where $\vec{x} \in \mathcal{X}^{n/k}$ represents the transmitted symbol sequence, and $\vec{y} \in \mathbb{C}^{n/k}$ denotes the received symbol sequence. The input alphabet is defined as $\mathcal{X} = \{x_1, x_2, \ldots, x_M\}$, a finite set of complex-valued constellation points, where each $x_m \in \mathbb{C}$ corresponds to a symbol in the modulation scheme. The output alphabet, denoted by $\mathcal{Y} = \mathbb{C}^{n/k}$, constitutes the continuous complex-valued space in which the received symbol sequence $\vec{y} \in \mathcal{Y}$ resides, capturing the effects of channel distortions and additive noise, as detailed below. This channel model is adaptable to various noise and fading scenarios prevalent in wireless communication systems.
Specifically, to enable comprehensive theoretical and numerical analysis of both conventional and NN-based transceivers, we focus on the additive white Gaussian noise (AWGN) channel. In this setting, the distortion and noise introduced by the channel can be modeled as:
\begin{align}\label{eq_awgn_channel}
\vec{y} = \vec{x}_{\text{norm}} + \vec{n}, \quad \vec{n} \sim \mathcal{CN}(\vec{0}, \sigma^2 \mathbf{I}),
\end{align}
where $\vec{x}_{\text{norm}} \in \mathcal{X}_{\text{norm}}^{n/k}$ is the normalized transmitted symbol sequence, and $\vec{n} \in \mathbb{C}^{n/k}$ is an AWGN vector.

\subsubsection*{Soft-Decision Decoding-Based Receiver}\label{sec_sdd_dm}

The input sequence to the receiver can be written as:
\begin{align}\label{eq_received_seq}
\vec{y} = (y^{(1)}, y^{(2)}, \ldots, y^{(n/k)}) \in \mathbb{C}^{n/k},
\end{align}
where $y^{(i)} = x_{\text{norm}}^{(i)} + n^{(i)}$, for $i = 1, \ldots, n/k$, and $n^{(i)} \sim \mathcal{CN}(0, \sigma^2), \forall i$ represents the AWGN component. The demodulator applies a mapping $g: \mathbb{C}^{n/k} \to \mathbb{R}^{n}$ to produce the estimate $\hat{\vec{c}}$ of the transmitted bit sequence $\vec{c}$.

In this paper, we consider a SDD-based receiver that employs APP-based LLR computation, as implemented in the Sionna library~\cite{sionna2022}. The demodulator processes the $i$-th received symbol $y^{(i)} \in \mathbb{C}$ to produce a vector of LLRs, $\vec{l}^{(i)} = [l^{(i)}(1), \ldots, l^{(i)}(k)] \in \mathbb{R}^k$, representing the confidence in the inference of each of the $k = \log_2 M$ coded bits associated with an $M$-ary constellation (e.g., for 16-quadrature amplitude modulation (16-QAM), $k = 4$).

Let $\vec{p}^{(i)} = [p_1^{(i)}, \ldots, p_k^{(i)}] \in \mathbb{R}^k$ denote the vector of prior LLRs for the $k$ bits of the $i$-th symbol, and let $b_{(i-1)k+j}$, for $j \in \{1, \ldots, k\}$, represent the $j$-th bit in the bit label $\vec{c}_i = [b_{(i-1)k+1}, \ldots, b_{ik}]$ of the transmitted symbol. The posterior probabilities of the $j$-th bit being 1 or 0, given the received symbol $y^{(i)}$ and prior LLRs $\vec{p}^{(i)}$, are denoted as $P(b_{(i-1)k+j} = 1 \mid y^{(i)}, \vec{p}^{(i)})$ and $P(b_{(i-1)k+j} = 0 \mid y^{(i)}, \vec{p}^{(i)})$, respectively.

Using Bayes' theorem, the LLR for the $j$-th bit of the $i$-th symbol is defined as:
\begin{align}\label{eq_llr_def}
l^{(i)}(j) = \log\left( \frac{P(b_{(i-1)k+j} = 1 \mid y^{(i)}, \vec{p}^{(i)})}{P(b_{(i-1)k+j} = 0 \mid y^{(i)}, \vec{p}^{(i)})} \right).
\end{align}

For an $M$-ary normalized constellation $\mathcal{X}_{\text{norm}}$, the constellation points are partitioned into subsets $\mathcal{X}_{j,1}$ and $\mathcal{X}_{j,0}$, containing the $M/2$ points where the $j$-th bit is 1 or 0, respectively. With Gray coding, adjacent constellation points differ by exactly one bit. Assuming an AWGN channel with noise $n^{(i)} \sim \mathcal{CN}(0, \sigma^2)$, where $\sigma^2 = N_0$, the LLR is computed as:
\begin{align}\label{eq_llr_awgn}
l^{(i)}(j) \!=\! \log \!\!\left(\! \frac{
\sum_{x \in \mathcal{X}_{j,1}} \!\! \Pr(x \mid \vec{p}^{(i)}) \!\! \exp \!\! \left( -\frac{1}{\sigma^2} \lvert y^{(i)} - x \rvert^2 \right)
}{
\sum_{x \in \mathcal{X}_{j,0}} \!\! \Pr(x \mid \vec{p}^{(i)}) \!\! \exp \!\! \left( -\frac{1}{\sigma^2} \lvert y^{(i)} - x \rvert^2 \right)
} \!\! \right) \!\!,
\end{align}
where the conditional probability (\textit{a priori} probability of $x$ given $\vec{p}^{(i)}$, with both having hypothetical roles) is:
\begin{align}\label{eq_prior_prob}
\Pr(x \mid \vec{p}^{(i)}) = \prod_{m=1}^k \text{sigmoid}(p_m^{(i)} \ell(x)_m),
\end{align}
and $\ell(x)_m \in \{-1, +1\}$ is the $m$-th bit label of constellation point $x$, with 0 mapped to $-1$ and 1 to $+1$. The sigmoid function is defined as $\text{sigmoid}(x) = \frac{1}{1 + e^{-x}}$.

When no prior knowledge is available ($\vec{p}^{(i)} = \vec{0}$), the prior probabilities are uniform ($\Pr(x) = \frac{1}{M}$), simplifying the LLR to:
\begin{align}\label{eq_llr_no_prior}
l^{(i)}(j) = \log\left( \frac{
\sum_{x \in \mathcal{X}_{j,1}} \exp\left( -\frac{1}{\sigma^2} \lvert y^{(i)} - x \rvert^2 \right)
}{
\sum_{x \in \mathcal{X}_{j,0}} \exp\left( -\frac{1}{\sigma^2} \lvert y^{(i)} - x \rvert^2 \right)
} \right).
\end{align}

The LLRs $\vec{l}^{(i)}$ are typically passed to a soft-decision decoder (e.g., turbo or LDPC) to recover the bit sequence. For hard-decision decoding (HDD), such as post-decoding, the estimated bit is determined by:
\begin{align}\label{eq_hard_decision}
\hat{b}_{(i-1)k+j} = 
\begin{cases}
1, & \text{if } l^{(i)}(j) > 0, \\
0, & \text{if } l^{(i)}(j) \leq 0.
\end{cases}
\end{align}

The final decoded bit sequence is $\hat{\vec{b}} = (\hat{b}_1, \hat{b}_2, \ldots, \hat{b}_n)$, where $\hat{b}_{(i-1)k+j}$ corresponds to the $j$-th bit of the $i$-th symbol. This sequence is compared to the transmitted bit sequence $\vec{b}$ to evaluate performance.

\subsection{Problem Formulation}

The APP-based SDD receiver, as implemented in the Sionna library~\cite{sionna2022}, computes LLRs for each bit of a received symbol to enable robust error correction in modern communication systems. However, its computational complexity, sensitivity to channel assumptions, and reliance on prior probabilities pose significant challenges, particularly for high-order modulation schemes. 
Specifically, for an $M$-ary constellation (e.g., 16-QAM with $M = 16$), the LLR for each of the $k = \log_2 M$ bits of a symbol $y^{(i)}$ is computed by summing likelihoods over $M/2$ constellation points in subsets $\mathcal{X}_{j,1}$ and $\mathcal{X}_{j,0}$, as described in Section~\ref{sec_sdd_dm}. The computational complexity of evaluating the likelihood terms $\exp\left( -\frac{1}{\sigma^2} \lvert y^{(i)} - x \rvert^2 \right)$ for each bit is $\mathcal{O}(M)$, leading to a total complexity of $\mathcal{O}(M \log_2 M)$ per symbol for all $k$ bits. For a sequence of $N = \frac{n}{k}$ symbols, where $n$ is the total number of bits, the overall complexity is $\mathcal{O}\left( N M \log_2 M \right)$.
This complexity scales unfavorably with $M$, posing a significant computational burden for high-order modulations (e.g., 256-QAM with $M = 256$).

The LLR computation incorporates prior probabilities through a vector of prior LLRs $\vec{p}^{(i)} = [p_1^{(i)}, \ldots, p_k^{(i)}] \in \mathbb{R}^k$, where $p_m^{(i)} = \log\left( \frac{P(b_{(i-1)k+m} = 1)}{P(b_{(i-1)k+m} = 0)} \right)$ for the $m$-th bit of the $i$-th symbol. When no prior information is available, a uniform distribution is assumed ($P(b_{(i-1)k+m} = 0) = P(b_{(i-1)k+m} = 1) = 0.5$), simplifying the prior probabilities to $\Pr(x) = \frac{1}{M}$ for each constellation point $x \in \mathcal{X}_{\text{norm}}$. In iterative decoding, non-uniform priors are derived from decoder feedback, but their accuracy depends on the decoding process and initial assumptions.

Moreover, the LLR computation assumes perfect knowledge of the noise variance $\sigma^2 = N_0$, where $n^{(i)} \sim \mathcal{CN}(0, \sigma^2)$ represents the AWGN for the $i$-th symbol. 
In contrast, symbol-level HDD, which selects the constellation point $x \in \mathcal{X}_{\text{norm}}$ minimizing $\lvert y^{(i)} - x \rvert^2$, has a lower complexity of $\mathcal{O}(M)$ per symbol, or $\mathcal{O}(N M)$ for $N$ symbols, as achieved by the proposed AE-CNN model presented later. However, HDD produces hard decisions without soft outputs, limiting its compatibility with modern error-correcting codes and resulting in higher BER compared to soft-decision decoding.

These limitations motivate our proposed approaches, i.e., NN-based systems that learn to approximate LLR computations with reduced complexity. Such models can infer effective prior probabilities from data, eliminate the need for explicit noise variance estimation, and achieve comparable or superior BER performance while addressing the computational limitations of conventional APP-based SDD receivers.

\section{Proposed NN-based Receiver and Transceivers}

Unlike the conventional techniques denoted in \ref{sec_sdd_dm}, NN approaches can achieve symbol detection through feedforward operations, offering less computational complexity and significant speed-ups in inference once trained. 
In this section, we propose \textit{NN-based bit-wise Demapper} that replaces the traditional LLR-based Demapper.
Furthermore, we extend the concept to an \textit{E2E Joint Optimal Symbol-wise AE-based Transceiver} where by jointly training the transmitter and receiver, the system can learns to map information symbols to robust representations that are effectively optimizing the communication system E2E.

\subsection{NN-based bit-wise Demapper}

\subsubsection{Forward Propagation the Proposed NN-based Demapper}

The Deep NN-based Receiver (DNN-Rx) is designed to learn a desired soft demodulation scheme, demapping a complex symbol to its conveyed bit's real-valued LLRs. Thus, the design of the NN-based Demapper is formulated as a regression problem,  where the model predicts continuous-valued LLRs corresponding to transmitted bits. The main objective is to minimize the discrepancy between the estimated probability and the ground-truth bit labels, obtaining comparable performance as APP-based SDD.

Considering the transmission of the sequence in \eqref{eq_vec_x_norm}.
At the receiver, a NN-based Demapper processes the sequence of complex-valued received symbols in \eqref{eq_received_seq}.
These symbols are first transformed into their real-valued representations and then are fed into NN model which outputs a sequence of $\vec{\hat{\ell}}=[\hat{\ell}_1, \hat{\ell}_2, \cdots ,\hat{\ell}_n ]$, corresponding to the estimated bit probabilities.

Each received $y^{(i)}$ is processed by the NN-based Demapper, modeled as a function $g_{\text{dm}} (\cdot; \vec{\theta}_{\text{dm}})$ where $\vec{\theta}_{\text{dm}}$ denotes the trainable parameters, producing a $k$-length vector LLRs. These output vectors are then concatenated to form a final output sequence $\vec{\ell}$, an $n$-length vector matching the transmitted bit sequence. The mapping from each received symbol to its LLR vector is modeled as:
\begin{align}
    \vec{\hat{l^{(i)}}} = g_{\text{dm}} (y^{(i)}; \vec{\theta}_{\text{dm}})
\end{align}

\subsubsection{Proposed NN Architecture}
The network includes three 
\textit{fully connected} (FF) layers. Before feeding to the network, complex symbol $y^{(i)}$ will be divided into its real and imaginary parts. The output vectors of the two input nodes are passed to the first hidden layer. Using composite function notation, e.g., $ (g \circ f) (x) = g (f(x))  $, a feedforward NN with three FF layers can be written as:
\begin{align}
\small
    g_{\text{dm}} = g^{[fc3]}_{\text{dm},\theta_{\text{dm}_3}} \circ g^{[a2]}_{\text{dm}} \circ g^{[fc2]}_{\text{dm},\theta_{\text{dm}_2}} \circ g^{[a1]}_{\text{dm}} \circ g^{[fc1]}_{\text{dm},\theta_{\text{dm}_1}}
\end{align}
and  $\vec{\theta}_{\text{dm}} =[\theta_{\text{dm}_1}, \theta_{\text{dm}_2}, \theta_{\text{dm}_3}]$  is the parameters of the NN-based Demapper. 
The output of $\ell$-th layer of it is:
\begin{align}
    g_{\text{dm}}\left({y}_{\ell-1}^{(i)};\theta_{\ell}\right) = \sigma\left(W_{\ell}  y_{\ell-1}^{(i)} +b_{\ell}\right),
    \label{DNN_layer}
\end{align}
where $W_{\ell} \in R^{N_{\ell} \times N_{\ell-1}}$ is called the weight and $b_{\ell} \in R^{N_{\ell}}$ are called the weight and bias, respectively ($R^{N_{l}}$ and $R^{N_{\ell-1}}$ are the dimension vector of $\ell$ layer and $\ell-1$ layer); the parameter of this layer is $\theta_{\ell} = [W_{\ell}, b_{\ell}]$; and $\sigma(.)$  is the activation function.
In our cases, the hidden layer uses a rectified linear unit (ReLU), namely $\sigma(z) = \max(0,y^{(i)}_{\ell})$ where $y^{(i)}_{\ell}$ is output of the $\ell$-th hidden layer. $\vec{l^{(i)}}$ is the desired output vector of the DNN-Rx. 
Before feeding into a loss function, for $j = 1, \dots, k$, the output of network passes through a sigmoid function \cite{2020Hoydis}:
\begin{align}
    \hat{l}^{(i)}(j) = \phi(l^{(i)}(j)) = 1 / (1 + \exp(-l^{(i)}(j))).
\end{align}
For each complex-valued input $y^{(i)}$, the output is a $k\times 1$ vector $\vec{\hat{l}}^{(i)} = [\hat{l}^{(i)}(1), ..., \hat{l}^{(i)}(j), ..., \hat{l}^{(i)}(k)] \in \mathbb{R}^k$.
\subsubsection{Loss Function Design and Back-Propagation}
The total output sequence of the NN-based Rx is an $n \times 1$ vector 
\begin{align}\label{eq_nn_dm_output}
    \vec{\hat{\ell}} = [\vec{\hat{l}}^{(1)},...,\vec{\hat{l}}^{(i)},..., \vec{\hat{l}}^{(n/k)} ] = [\hat{\ell}_1, \hat{\ell}_2, \cdots ,\hat{\ell}_n ].
\end{align}
The model is trained by minimizing the Binary Cross-Entropy (BCE) as the objective is to minimize the BER.
Note that $b_i$ is the ground-truth bit, which serves as the true hypothesis. If the bit sequence has length $n$ and the batch size is 1, then we compute the BCE loss by first passing each estimated LLR through the sigmoid function to obtain the predicted probability. The loss is then calculated by comparing $\hat{\ell}_{i}$ with $b_i$ for all $i = 1, \dots, n$, i.e.:
\begin{align} \label{eq_loss_dnn_dm}
    \textstyle \mathcal{L}_{\text{dm}} \!=\! - \frac{1}{n}\sum_{i=1}^{n} \left[ b_i \log \hat{\ell}_{i} \!+\! (1 \!-\! b_i) \log \bigl(1 \!-\! \hat{\ell}_{i}\bigr) \right],
\end{align}
The loss $\mathcal{L}_{\text{dm}}$ is minimized by computing the gradient of the BCE with respect to the neural demapper’s parameters $\vec{\theta}_{{dm}}$ using backpropagation. These parameters are then updated via optimizer Adam, which adjusts them in the direction that reduces $\mathcal{L}_{\text{dm}}$. Then $\vec{\theta}_{{dm}}$ is updated after one iteration. The iteration continues until the loss converges.

\subsubsection{Training Process} 

 The training procedure begins with the generation of random binary sequences of batch size $\mathcal{B}$ sequence of bits. During training, the model parameters are initialized and optimized at a fixed energy-per-bit to noise power spectral density ratio $\mathcal{E}_b/N_0$. The loss function is computed as in \eqref{eq_loss_dnn_dm}, and the model parameters are updated through iterative optimization to minimize this loss. The performance of the NN-based Demapper is evaluated using Monte Carlo method across a range of $\mathcal{E}_b/N_0$ values. Before comparison with the ground-truth bits, the predicted LLRs are passed through a hard-decision thresholding process as in \eqref{eq_hard_decision}.

\subsection{Proposed Symbol-wise AE-based E2E Transceiver}

In the E2E architecture based on AEs has been proposed, 
the transmitter, the channel, and the receiver are modeled as an encoder, a bottleneck layer, and a decoder, respectively. We propose symbol-wise approach with the primary objective of minimizing the symbol error rate (SER). 

\subsubsection{Forward Propagation of the AE-based E2E Model}

Considering the principle of MAP, our proposed communication system is treated as a \textit{classification problem} for which the Categorical Cross-Entropy (CCE) is the loss function. The AE-based communication system is typically trained by minimizing the CCE between the posterior distribution and the one learned by the model.
In the forward propagation, the single bit-block $\vec{c_i}$, representing $k$ information bits, as in \eqref{eq_bit_block}, is first mapped to an integer index $m$ corresponding to its position in the codebook of size $M=2^k$. This index $m$ is then encoded as a one-hot vector, which serves as the input to the encoder network (EncNet). The encoder transforms this input into a modulated symbol, which is a NN-based version of \eqref{eq_vec_x_norm}, and then is transmitted over the channel, i.e., the bottle-neck layer. The received noisy symbol, i.e., the NN-based version of \eqref{eq_received_seq}, is subsequently passed through the decoder network (DecNet), which outputs an $M-$dimensional probability vector. 
The E2E communication is treated as a single trainable function as:
\begin{align}
     g_{\text{e2e}, \vec{\theta}_{\text{e2e}}} = g_{\text{rx}\vec{\theta}_D} \circ g_{\text{ch}} \circ g_{\text{tx},\vec{\theta}_M}
\end{align}
where  $g_{\text{tx}}, g_{\text{ch}}, g_{\text{rx}}$ represents as transmitter, channel, and receiver function, respectively, with parameter $\vec{\theta}_{\text{e2e}} =\{\vec{\theta}_D, \vec{\theta}_M\}$.

\subsubsection{AE-CNN Architecture}

In our analysis (which is omitted here), the AE-CNN achieves lower computational complexity compared to AE-DNN. Assume the channel output size of all convolution layers is $C_{\text{out}}$, AE-CNN exhibits a complexity of $\mathcal{O}(2^kNC_{\text{out}})$, whereas AE-DNN has a complexity of $\mathcal{O}(2^{2k}N)$. As a result, the computational complexity of AE-DNN will grow exponentially with $k$ which highlights the substantial gap in computational efficiency between the two architectures.

\paragraph{EncNet for the Transmitter}
Considering $\vec{c}_i$ in \eqref{eq_bit_block}, the transmitter of the proposed E2E-AE maps $\vec{c}$ to complex baseband vector $\vec{z} $. 
We denote the transmitter by the function
\begin{align}
g_{\text{tx},\vec{\theta}_M} : \{0,1\}^k \,\rightarrow\, \mathcal{Z}_{\text{norm}} \in \mathbb{C},
\end{align}
with learnable parameters $\vec{\theta}_M =[\theta_{M1}, \theta_{M2}, \theta_{M_3}]$, 

The system follows by the transformation of $\vec{c}$ messages into one-hot vector $\vec{1}_m$. Since there are $M$ possible messages, each $\vec{c_i}, i= 1,\dots,n/k$ is typically encoded as a $i$-th one-hot vector $\vec{1}_{m} \in \mathbb{R}^M$, $m \in 1,2, \dots,M$. This step expands a $k$ bits input into one of $M = 2^k$ possible one-hot vectors. Let $\vec{1}_{i,m}(r) = 1$ if $r = m$ and $0$, otherwise, where $r$ is the index position in the one-hot vector $\vec{1}_{i,m}$.
The one-hot vector $\vec{1}_{i,m}$ is then passed through EncNet, which is described as:
\begin{align}
   g_{\text{tx},\theta_M}\left(\vec{1}_{i,m}(r)\right)
   \;=\;
   g^{\text{norm}}_{\text{tx}}
   \circ
   g^{\text{R} \to \text{C}}_{\text{tx}}
   \circ
    g^{[a_3]}_{\text{tx}}
    \circ \\
  g_{\text{tx},\theta_{M3}}^{[fc_3]}
   \circ
   g^{[a_2]}_{\text{tx}} \circ  g_{\text{tx},\theta_{M2}}^{[cnn_2]}
   \circ
   g^{[a_1]}_{\text{tx}} \circ  g_{\text{tx},\theta_{M1}}^{[cnn_1]}
   \left(\vec{1}_{i,m}(r)\right),\nonumber
\end{align}
where $g_{\text{tx},\theta_{Mi}}^{[\cdot]}$ denotes a learnable NN layers (either 1D-CNN or DNN) with trainable parameters $\theta_{Mi}$, and each $g_{\text{tx},\theta}^{[a_i]}$ denotes a non-linear activation applied after the corresponding layer, $g^{\text{R} \to \text{C}}_{\text{tx}}$ converts the real-value output into a complex-valued representation. Finally, $g^{\text{norm}}_{\text{tx}}$ enforces power constraints  $
\mathbb{E}[|\vec{x}^{(i)}|^2] \le 1$. The output at the $\ell$-th convolutional layer, resulting from the  function  $g^{[cnn]}_{\text{tx}}$, can be expressed as: \begin{align} 
g_{\text{tx}}\left(y_{\ell-1}^{(i)}, \theta_{\ell}\right) = w_{\ell} \circledast y_{\ell-1}^{(i)} + b_{\ell}, \label{CNN_Layer}
\end{align} 
where $w_{\ell}$ denotes the set of convolutional filters (kernels) in the $\ell$-th layer, and $b_{\ell}$ is the corresponding bias term. The trainable parameters of the $\ell$-th convolutional layer are given by $\theta_{\ell} = [w_{\ell}, b_{\ell}]$. The operator $\circledast$ indicates the convolution, with each filter in $w_{\ell}$ sliding over the input feature map $y_{\ell-1}^{(i)}$ to extract local features. The transmit signal is expressed as
$
z^{(i)}_{\text{norm}}
\;=\;
g_{\text{tx},\theta_M}(\vec{c}_{i})
\;\;\in\;\;
\mathbb{C}.
$

\subsubsection{Bottleneck Layer for the Channel}

The channel is modeled as a noisy bottleneck in the AE by adding redundancy. The purpose of AE is to learn representations $\vec{z}$ of the message $\vec{c}$ that is robust with respect to the channel impairments mapping $\vec{z}_{\text{norm}}$ to $\vec{y}$ so that the transmitted message can be recovered with small probability of error. In AWGN channel model, the received signal $\vec{y}$ can be represented as \eqref{eq_awgn_channel}. From an AE standpoint, this channel layer is inserted between the EncNet (transmitter) and the DecNet (receiver) to emulate real physical impairments.

\paragraph{DecNet as Receiver}
The receiver is a NN that attempts to reconstruct the original message $\vec{c}$ from the noisy observation $\vec{y}$. Denote this decoder by
\begin{align}
g_{\text{rx}\theta_D} : \mathbb{C} \,\rightarrow\, \mathbb{R}^k,
\end{align}
with learnable parameters $\vec{\theta}_D =[\theta_{D1}, \theta_{D2},\theta_{D_3}]$. Typically, the receiver network also comprises several layers to perform detection and symbol demapping:
\begin{align}
g_{\text{rx},\theta_D}\bigl(y^{(i)}\bigr)
\;=\;
g_{\text{rx},\theta_{D3}}^{[a_3]} \circ g_{\text{rx},\theta_{D3}}^{[fc_3]}
\circ\\ \nonumber
g_{\text{rx},\theta_{D2}}^{[a_2]} \circ g_{\text{rx},\theta_{D2}}^{[cnn_2]}
\circ
g_{\text{rx},\theta_{D1}}^{[a_1]} \circ g_{\text{rx},\theta_{D1}}^{[cnn_1]}
\circ g^{\text{C} \to \text{R}}_{\text{rx}}({y}^{(i)}),
\end{align}
where $g_{\text{rx}}^{\text{C} \to \text{R}}$ will convert complex inputs into real and imaginary parts for subsequent layers. While  $g_{\text{rx},\theta_{D2}}^{[cnn_2]}$ and $g_{\text{rx}\theta_{D1}}^{[nn_1]}$ 
can be calculated similarly as in \eqref{CNN_Layer}, the  $g_{\text{rx}\theta_{D3}}^{[fc_3]}$ denotes the fully connected layer as in \eqref{DNN_layer}.  $g_{\text{rx}\theta}^{[a_i]}$ in receiver also denotes the activation function of corresponding layer.
In terms of \textit{Symbol-wise AE}, the decoder produces $\vec{p} \in(0,1)^M$ which is a posterior distribution probability vector over all possible messages. The softmax activation function is applied at the last layer of the DecNet, which can be expressed as:
\begin{align}
    \vec{p}(i) = \phi(y_{\text{last layer}}^{(i)}).
\end{align}

\subsubsection{Loss Function}

The AE-based E2E transceiver is optimized on the symbol-wise CCE. The optimization objective is to minimize the CCE loss between the true message label $\vec{1}_m$ and the predicted probability vector $\vec{p}$ at the receiver output
\begin{align}
        \textstyle \mathcal{L}(\theta_{M},\theta_{D}) =  - \frac{1}{n/k}\sum_{i}^{n/k} \vec{1}_{i,m}(r) \log \vec{p}(i).
\end{align}

\subsubsection{Training and Inference}

The model is trained at fixed values of the energy-per-bit to noise power spectral density ratio, $\mathcal{E}_b/N_0$ using Adam optimizer, and evaluated at a wide range of $\mathcal{E}_b/N_0$ values. During the training phase, we minimize the CCE loss between target encoded one-hot vector $\vec{1}_m$ and predicted probability $\vec{p}$. After each iteration, the parameters of both the transmitter and receiver are updated accordingly. To identify the robust model, we conduct training experiments at various $\mathcal{E}_b/N_0$. During the inference, the model outputs the posterior probability with the highest probability at $r$-index position which then maps to the estimated symbol $\hat{m}$, i.e., $\hat{m}_i = \arg\max_{r}\left\{\vec{p}(i)\right\}$.
The estimated symbol vector $\vec{\hat{m}}$ is then compared to the truth index $\vec{m}$ to compute the SER. 

\section{Performance Analysis and Numerical Results}\label{section:result}

\subsection{Error Probability Analysis}

We assess the performance of the examined information-theoretic systems in terms of SER and BER. The design of NN-based systems and their loss functions is selected depending on whether the goal is to minimize BER or SER. Specifically, the AE-based E2E system evaluated for SER is optimized using symbol-wise CCE, whereas the system evaluated for BER is optimized using BCE.

\begin{remark}\label{theorem_SER_approx}
The approximate closed-form expression for the SER of square $M$-QAM is derived as follows:
\begin{align}
   \mathrm{SER} \approx \frac{4(\sqrt{M}-1)}{\sqrt{M}}   Q\left(\sqrt{\frac{3P_{s}B}{\log_2M(M-1)N_0R}}\right).
   \label{eq_final_ser}
\end{align}
\end{remark}
Owing to space constraints, we exclude the derivation of the aforementioned closed-form expression. 

For BER analysis, to determine the probability that the $i$-th bit in a symbol is in error, there are $2^{k-1}$ symbols with the same bit value and $2^{k-1}$ symbols with the opposite bit value. This implies that $2^{k-1}$ symbol error cases result in this particular bit being altered. Recall that, in Gray Coding rule, if $i$-th bit in that symbol error then which symbols also have effect. The BER is derived in the following remark.

\begin{remark}
For square $M$-QAM, using \eqref{eq_final_ser}, the closed-form expression for BER is obtained as:
\begin{align}\label{eq_final_ber}
\mathrm{BER} = \frac{2^{k-1}}{2^k - 1} P_e = \frac{M/2}{M-1} \mathrm{SER}.
\end{align}
\end{remark}

\subsection{Results and Numerical Analysis}

\begin{figure}[t]
    \centering
    \includegraphics[width=.9\linewidth]
    {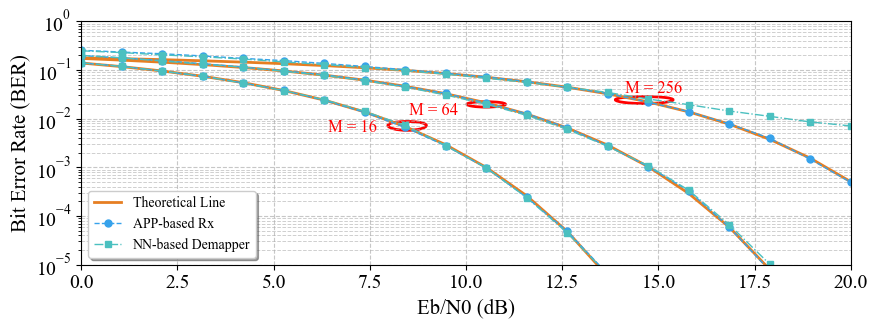}
    \caption{BER performance of NN-based bit-wise demapper (Rx), APP-based demapper, and theoretical results for various modulation orders.}
    \label{fig_NNDemapper}
\end{figure}

\begin{figure}[t]
    \centering
        \includegraphics[width=.9\linewidth]{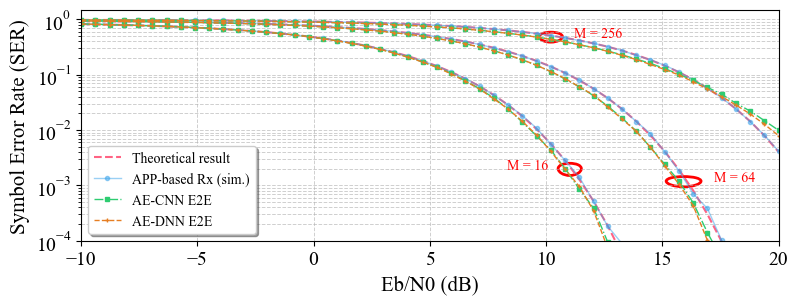}
    \caption{SER performance of symbol-wise AE-based transceivers (AE-DNN and AE-CNN), APP-based system, and theoretical results in \eqref{eq_final_ser}.}
    \label{fig_AE_CNN_DNN}
\end{figure}

The training settings, NN architecture configuration, and wireless-related parameters are detailed in our published code repository and are therefore omitted here to save space.
First, we validate the correctness of our NN-based results. As observed in Figs.~\ref{fig_NNDemapper} and \ref{fig_AE_CNN_DNN}, the numerical NN-based results are well corroborated by theoretical and Monte Carlo simulation results, confirming that NN-based systems can operate as digital twins of conventional information-theoretic systems.
In Fig.~\ref{fig_NNDemapper}, we present a Monte Carlo simulation of a conventional APP-based SDD receiver alongside the theoretical and NN-based results. As shown in Fig.~\ref{fig_NNDemapper}, the three methods yield similar performance in low SNR regimes and with low and intermediate modulation orders, e.g., $M \leq 64$.

In Fig.~\ref{fig_AE_CNN_DNN}, for performance comparison, we implement an additional AE-DNN, where the EncNet and DecNet consist of FF layers. As observed in Fig.~\ref{fig_AE_CNN_DNN}, the NN-based systems outperform the conventional system at low SNR and maintain comparable performance at high SNR with specific high $M$. The AE-CNN yields results similar to the AE-DNN at certain ${\mathcal{E}_b}/{N_0}$ values, consistently surpassing the conventional baseline at low $M$. It is noted that during the training process, the AE-DNN required more time to converge than the AE-CNN, likely due to its higher computational complexity, as discussed in the problem formulation.

\begin{figure}[t]
    \centering{%
        \includegraphics[width=.9\linewidth]{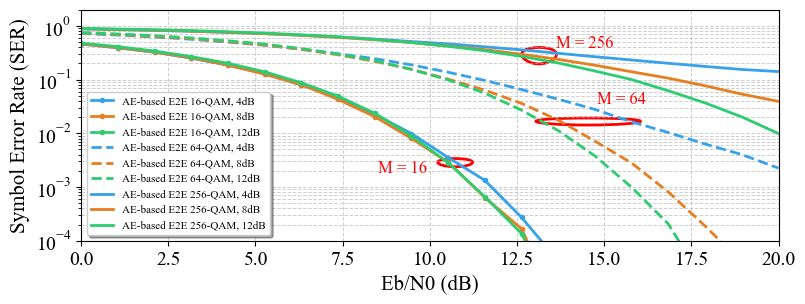}
    }
    \caption{Trained AE-CNN performance under different inference conditions.}
    \label{fig_AECNN_differentSNR}
\end{figure}

Finally, we seek to determine the most robust model among all proposed NN-based systems, as training at low SNR levels does not ensure the capture of structural features critical for optimal performance at higher SNR conditions. In Fig.~\ref{fig_AECNN_differentSNR}, the AE-based E2E model trained at $8$ dB demonstrates superior performance at low ${\mathcal{E}_b}/{N_0}$, while the model trained at $12$ dB outperforms it in high SNR regions across all three modulation schemes. Our numerical study, detailed in our GitHub repository, shows that the AE-based transceiver exhibits greater robustness than the NN-based demapper in terms of inference performance across varying SNR levels.

\section{Conclusion}
In this paper, we introduced NN-based information-theoretic solutions in physical layer where the conventional APP-based SDD methods suffer high computational complexity, particularly with high-order modulations. In contrast, our proposed NN-based Receiver are trained directly on the data achieving superior BER performance under AWGN channel. We further extend our approach to AE-CNN-based E2E transceiver, which outperforms traditional methods. Additionally,  we also demonstrate the relationship between training conditions and the robustness of the model during inference.
\bibliographystyle{IEEEtran}
\bibliography{References}

\end{document}